# Hasq Hash Chains

Oleg Mazonka and Vlad Popov

Hasq Technology Pty Ltd, Australia, 2014
om@hasq.org, vp@hasq.org

**Abstract.** This paper describes a particular hash-based records linking chain scheme. This scheme is simple conceptually and easy to implement in software. It allows for a simple and secure way to transfer ownership of digital objects between peers.

## Contents

## 1    Introduction

The concept of ownership implies control. This is true for real objects as well as for abstract concepts like a bank account, copyright or digital data. The examples of owner's actions controlling their assets include transferring ownership to another party at will. However, in reality one's ability to control is somewhat limited due to a fundamental property of most things that can be owned - ownership can be lost against will. For example, cash can be lost as with other physical objects, a bank account can be locked, copyright may be lost due to a law suit, digital data can be unlawfully copied, effectively destroying its owner's control.

There is a concept, however, which is substantially different from the above and that is authorship. Authorship cannot be taken away and it cannot be transferred. Table 1 summarizes these findings.

While in some cases the loss of control may be a result of a legitimate decision, in others it may be the product of criminal actions and, as such, is highly undesirable. Over the years substantial resources have been devoted to developing solutions both for physical and electronic security. Digital data is particularly vulnerable due to the ease of copying. At the same time the importance of protecting such data is increasing as information technology plays a greater and more critical role in modern society.

A natural solution to the digital data ownership problem is an electronic register that keeps track of records associating owners and their digital objects. In order to use such a register, owners must place trust in it which is not always desirable. Also, operational overheads often add costs to the use of such registers and this may limit their customer base. There is a long history of attempts to relax the requirement to trust a register and reduce the costs [1-6]. One of the latest breakthroughs in this area is a public ledger blockchain technology, originally implemented in Bitcoin cryptocurrency [7].

Two ownership properties shown in Table 1 follow a Yes-Yes pattern for usual objects and No-No for authorship. An object with a No-Yes pattern, however, would correspond to something that has the same intrinsic protection from being taken away (as with authorship), but with the ability to be transferred if its owner wishes so. An object of this kind would have a multitude of applications since it would make practical use of the idea of ultimate protection.

In the following section we present a scheme that allows one to control a digital object (Hasq token) as secure - as if the owner is the author of a publicly available artifact - and at the same time being able to transfer ownership to another untrusted party.

| Ownership property | Real objects, money (cash) | Money (bank account) | Copyright | Authorship | Hasq token |
|---|---|---|---|---|---|
| Can be lost against will | Yes | Yes | Yes | No | No |
| Can be transferred at will | Yes | Yes | Yes | No | Yes |

Table 1

## 2 Hash Chain

Suppose there is a publicly available textual database. One can think of a database as an equivalent of issues of a magazine or a newspaper. The database is built upon a particular hash function and consists of a list of records. Each record has the following textual fields separated by a space:

$$N \quad S \quad K \quad \{G_1 \quad G_2 \quad ...\} \quad O \quad [D]$$

where
$N$ is a sequence number for a particular $S$;
$S$ is a hash function representation of digital data of arbitrary nature, $S$ is expressed as a string of hexadecimal[1] characters (*token*);
$K$, $G$ and $O$ are hash strings explained below;
$D$ is an optional textual data field.

The fields $K$, $G$ and $O$ are used to link records. They stand for Key, Generator and Owner. A number of fields $G$ is arbitrary but fixed within a database. For the sake of simplicity we will consider only one field $G$; however, zero or more than one can easily be assumed by a reader. A database is a mix of records containing different tokens $S$. If records with the same $S$ are selected, the list of records would look like the following (the field $D$ is left out here for simplicity since it does not participate in record linking):

...
$N_0 \quad S \quad K_0 \quad G_0 \quad O_0$
$N_1 \quad S \quad K_1 \quad G_1 \quad O_1$
$N_2 \quad S \quad K_2 \quad G_2 \quad O_2$
$N_3 \quad S \quad K_3 \quad G_3 \quad O_3$
$N_4 \quad S \quad K_4 \quad G_4 \quad O_4$
...

Here $N_0, N_1, N_2...$ are consecutive numbers for the chosen $S$: $N_1 = N_0 + 1$, $N_2 = N_1 + 1$, and so on.

The records are chained with use of the following rules:

---

[1] Here we describe a particular selection of forms such as textual representation or concatenation character. This is specified explicitly for ease of presentation; however, some generality is sacrificed.

$$G_0 = \text{Hash}(N_1, S, K_1)$$
$$O_0 = \text{Hash}(N_1, S, G_1)$$
$$G_1 = \text{Hash}(N_2, S, K_2)$$
$$O_1 = \text{Hash}(N_2, S, G_2)$$
and so on

Hash function takes arguments in a textual representation with arguments concatenated with the use of a space character. The result is also a textual representation in hexadecimal format.

A server hosting the database publishes a new record *only* if it follows the above rules. Such a database is said to be consistent and its consistency can be verified independently.

The following illustration is a graphical representation of the above rules.

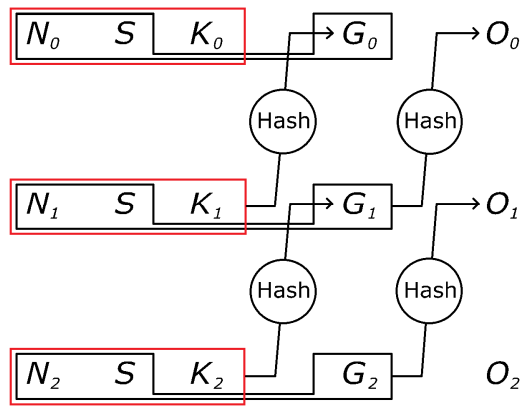

Figure 1

User ownership is realized in the following way. Suppose that the last record the database contains for a particular $S$ is $N_0$ $S$ $K_0$ $G_0$ $O_0$. It is said that the user owns token $S$ when he knows the (secret) keys $K_1$ and $K_2$. These keys in turn can be generated by some algorithm from the user's personal *passphrase*. For example, a simple key generation method could be:

$$K_i = \text{Hash}(i, S, \text{passphrase})$$

There are number of techniques to pass token ownership from one person to another. The most interesting scenario is when the recipient (who receives the token – the future owner) desires to stay absolutely anonymous[2]. The recipient generates $K_3$, $G_2$ and $O_1$, then sends $O_1$ to the current owner, let us call him the sender. The sender publishes a new record $N_1$ $S$ $K_1$ $G_1$ $O_1$ revealing his first secret key $K_1$. After the publication the recipient can verify that $O_1$ appeared in the database. At this moment in time neither the sender nor the recipient owns the token $S$, because $K_2$ is known only to the sender and $K_3$ is known only to the recipient. Until the sender and the recipient come to an agreement, the token is locked. Next the recipient generates $O_2$ (by generating $K_4$ and $G_3$ first) and sends both $G_2$ and $O_2$ to the sender to publish the record $N_2$ $S$ $K_2$ $G_2$ $O_2$. Once this record is published, the recipient becomes the current owner of token $S$ because he knows both new secret keys $K_3$ and $K_4$.

---

[2] This distinguishes from Bitcoin's pseudo-anonymity. Publication of records is based on messages holding records passed to a database keeping server. Generally servers do not require TCP or other real-time connection with a client.

## 3 Technology and Applications

Hasq Technology Pty Ltd has developed a working high performance distributed system which utilizes the above described scheme and has solved many issues accompanying the implementation.

Hasq databases are shared between untrusting servers by a simple network protocol. The core of the protocol is a set of commands to get records and to add a new record. The servers are tuned to send notifications to other servers when they publish new records, but their work is independent and the databases are not required to be synchronized.

Isolated failures of servers or connection issues do not affect the stability of the Hasq network as a whole since servers are designed to reconfigure themselves to maintain network integrity as much as possible.

To accommodate different hardware requirements several types of servers are supported. Effective control of server database size is also possible via use of configuration parameters.

It may be desirable in some cases to make databases incompatible. Different hash functions help achieve this. If the same hash function is used, special fixed publicly known strings added to all hash calculations make databases unique.

The Hasq record chaining scheme allows manipulation of tokens with the following properties. The token can:

- be passed electronically over the Internet or telephone
- be passed fully anonymously for either the sender or the recipient
- serve as a currency note
- serve as an obligation
- serve as a verifiable title for ownership
- serve as modifiable association with a particular piece of electronic data

If an organization issuing tokens sets up a group of non-anonymous users, the organization may use some tokens in the form of a debt, i.e. tokens representing negative values. This can be organized as follows. Each user owns an ID token representing their identity. User *A* having a particular predefined text (debt token), published in their data field, is assumed to owe something. Once another user *B* is willing to accept this debt from user *A*, user *B* publishes the debt token in the data field of their identity token. User *A* publishes a reference to the record where user *B* accepted the debt token, effectively releasing himself from the debt token.

The above is a short example of many other applications [8] already known to the community. This list may grow over time considering current technologies are in their early stages.